\documentclass[referee]{raa}            

\usepackage{graphicx,times}             
\usepackage{hyperref}
\usepackage{filecontents}
\usepackage{amssymb}
\usepackage{xcolor}
\usepackage{natbib}
\usepackage{amsmath}

\newcommand{\gx}{GX~339-4}

\begin{document}

   \title{Systematic Analysis of Low/Hard State RXTE Spectra of GX 339-4 to Constrain the Geometry of the System
}

   \volnopage{Vol.0 (200x) No.0, 000--000}      
   \setcounter{page}{1}          

   \author{Kalyani Bagri
      \inst{1}
   \and Ranjeev Misra
      \inst{2}
   \and Anjali Rao
      \inst{2}
   \and J. S. Yadav
      \inst{3}
    \and S. K. Pandey
      \inst{1} 
   }

   \institute{Pt. Ravishankar Shukla University, Raipur,
              India; {\it kalyanibagri@gmail.com}\\
        \and
             Inter University Center for Astronomy and Astrophysics, Pune\\
        \and
             Tata Institute of Fundamental Research, Mumbai\\
   }

   \date{Received~~2016 month day; accepted~~20016~~month day}

\abstract{ One of the popular models for the low/hard state of Black Hole Binaries is that the standard accretion disk is truncated and the hot
inner region produces via Comptonization, the hard X-ray flux. This is supported by the value of the high energy photon index, which is
often found to be small $\sim$ 1.7 ($<$ 2) implying that the hot medium is seed photons starved. 
On the other hand, the suggestive presence of a broad relativistic Fe line during the hard state would suggest that the accretion disk is not 
truncated but extends all the way to
the inner most stable circle orbit.
In such a case, it is a puzzle why the hot medium would remain photon starved. The broad Fe line should be
accompanied by a broad smeared reflection hump at $\sim$ 30 keV and it may be that this additional component makes the spectrum hard
and the intrinsic photon index is larger, i.e. $>$ 2. This would mean that the medium is not photon deficient, reconciling the presence of a
broad Fe line in the observed hard state. To test this hypothesis, we have analyzed the RXTE observations of GX 339-4 
from the four outbursts during 2002-2011
and identify the observations when the system was in the hard state and showed a broad Fe line. We have then attempted to fit
these observations with models, which include smeared reflection to understand whether the intrinsic photon index can indeed be
large. 
We find that, while for some observations the inclusion of reflection does increase
the photon index, there are hard state observations with broad Fe line that have
photon indices less than 2.
\keywords{accretion, accretion disks -- stars: black holes -- X-rays: binaries -- X-rays: individual (GX 339-4)}
}

   \authorrunning{Bagri et al.}            
   \titlerunning{Systematic Analysis of Low/Hard State RXTE Spectra of GX 339-4}  

   \maketitle
\section{Introduction}           
\label{sect:intro}
Black hole binaries are some of the highly variable astronomical 
objects observed in the X-ray sky
and their variability is revealed in both spectral and timing characteristics.
The occasional presence of Quasi Periodic Oscillations (QPOs), variation of their 
centroid frequency, and rare occurrence of high frequency QPOs etc. are some of the manifestations
of variability observed in timing properties. 
The variability in spectral behavior can be described in terms of spectral states exhibiting
different shapes, which
can be interpreted as being due to varying relative contribution from two or more spectral 
components. The two major components observed in the spectrum include a multi-temperature
thermal component (\cite{mitsuda84, shakura73}) originating from 
an accretion disk and a power law tail due to
inverse Comptonized radiation from the corona. 
The thermal component is mainly present towards 
softer X-ray bands ($\leq$10 keV)
and it dominates the overall spectrum when a system is in the soft spectral states.
The copious amount of soft seed photons originating from the disk results in a steep 
spectrum with a photon index $\Gamma \geq$2. However, the strength of the thermal
emission from the disk weakens during the hard spectral state and the spectrum
is dominated by a hard power law component that extends towards the softer energies
as well. The weak thermal component implies
a smaller number of soft seed photons resulting in a relatively flat spectrum with 
a photon index of $\Gamma <$2. The transient black 
hole binaries undergo a series of spectral state transitions during an outburst in a 
systematic manner, which is often depicted as Q-shaped track 
in hardness-intensity diagrams \citep{fender09}. These 
diagrams have shown that a transient black hole binary is observed in the hard 
state when it enters the outburst, followed by transition to the hard intermediate state (HIMS),
soft intermediate state (SIMS) and soft state \citep{belloni11,fender09,homanbelloni05}.
The order is reversed as the outburst begins to decline and a system may show a few
excursions to different spectral states \citep[see][]{belloni11,fender09,homanbelloni05}. 
However, the overall scheme of state transitions
remains the same for transient black hole binaries. On the other hand, the persistent
black hole binaries have not shown a predictable trend of state transitions as in the 
case of transient black hole binaries, for example, Cyg X-1 remains in the hard state most of the 
time with occasional excursions to the soft state \citep{grinberg}. 

The seed photon starved hard spectral state is generally understood in terms of an accretion 
disk truncated far from the central black 
hole and the inner region being replaced by the hot inner flow which is sometimes modeled as an 
Advection Dominated Accretion Flow \citep[ADAF; ][]{narayanyi95}. The truncation of disk
may be described by the disk evaporation model, which was first proposed by \cite{meyer2K}
and later extended by \cite{liu02} and \cite{meyermeyer03}. The model
considers a continuous evaporation of material from the disk, which feeds the corona. 
\cite{qiaoliu09} calculated the evaporation rate as a function of disk radius and showed that
it is maximum at a certain radius for a given value of the viscosity 
parameter. The model predicts that the truncation of disk will occur at a radius where the 
accretion rate is equal to the maximum evaporation rate and the inner disk will 
survive the evaporation only when the accretion rate is higher
than the maximum evaporation rate. 
The model successfully explains the finding of a truncated accretion disk 
\citep[e.g.][etc.]{plant14,plant15,tomsick09} as a consequence of the low mass 
accretion rate during the hard state, and an extended disk approaching the 
Innermost Stable Circular Orbit (ISCO) 
owing to the higher mass accretion rate in the soft state. It also provides 
a framework to the models addressing the steady jet emission during the hard
states (see \cite{belloni10} and \cite{done07} for detailed reviews).

The estimation of the inner disk radius using the observational data is 
crucial to understanding the truncation
of accretion disk, which can be measured by modeling of Fe line profile and disk continuum.
If the accretion disk is truncated during the hard states, the Fe 
emission line should be narrow and symmetric as the relativistic
effects due to gravitational field of the black hole are weaker on the disk material.
However, in soft states, the line should be broadened and skewed by the strong 
general relativistic effects in the inner region. Therefore, finding a broad Fe line in hard 
state serves as a strong evidence towards the extension of the disk in the 
inner regions and hence the
violation of the truncated disk model. 
In such a scenario, a seed photon starved system during the hard state remains a puzzle. In several 
reports, authors have not only found a broad and skewed Fe line in the hard state, but also 
estimated the spin parameter by modeling the line profile and found a value of spin parameter consistent
with the results obtained from the study of continuum and reflection component. 
\cite{miniutti04} studied the three BeppoSAX observations of a black hole
candidate XTE J1650-500 during its 2001-02 outburst and perhaps detected the 
source in the power law dominated hard state ($\Gamma \sim$1.8) in one of the observations.
They find a broad and strongly relativistic Fe emission line in the spectrum. 
The modeling of the line revealed the presence of the disk extending to $\sim$1.34 $R_g$, 
which led them to suggest the presence of a Kerr black hole in the system.
The extension of the accretion disk to the inner region in another black hole 
binary Swift J1753.5-0127 is shown by \cite{miller06} during the decline 
of its 2005 outburst, wherein the inner disk was found at or close to 
the ISCO by modeling the continuum with a number of models. Their results
show that the disk can be present in the inner regions during the hard state
at very low luminosities down to $L_{X} \simeq$ 0.003 $L_{Edd}$.
The same set of observations belonging to the hard state were further studied
by \cite{reis09-395} to find the spin parameter and inclination angle using a
model that included a power law and reflection component. The innermost emitting
region was shown to extend close to $\sim$3.1$R_g$ and a spin parameter of $\sim$0.76 
was reported by assuming that the inner edge of the disk is at the ISCO.
Another result presenting the violation of the truncated disk model is shown in
\cite{reis09l52} wherein the authors extracted the spectrum of XTE J1118+480 in its 
hard state using Chandra and RXTE and found the presence of thermal 
component with disk temperature of $\sim$ 0.21 keV, which suggested
that the emission is originating from an accretion disk extending close to the radius
of marginal stability. 
XTE J1817-330 was studied by \cite{rykoff07} using Swift observations
during its 2006 outburst. They used the disk continuum model and found
the inner disk radius to be consistent with the ISCO. They showed that the 
luminosity follows the relation $L_X \propto T^4$ roughly during 
the decline of the outburst, which led them to suggest the presence of a geometrically
stable disk in the inner regions at accretion rate as low as 0.001 $L_{Edd}$.
\cite{reis10} studied eight black hole binaries in their hard spectral
state. The modeling of the disk continuum revealed that the luminosity in all 
the systems is consistent with the relation $L_X \propto T^4$ down to $\sim$
5$\times$10$^{-4} L_{Edd}$. The six sources showed truncation radius 
not larger than 10 $R_g$ and the Fe line detected in four of 
the black hole binaries at luminosities down to 1.5$\times$10$^{-3} L_{Edd}$
excluded a truncated disk.
Other studies on the finding of the inner disk radius close to the ISCO during
low/hard state include \cite{reynolds10} and \cite{reynoldsmiller13}. Hence, there are 
good number of
evidence suggesting that the disk may not be truncated in the hard spectral state.

While several results show the violation of the truncation model, there are also 
number of results that support the truncation of the disk in the hard state.
\cite{tomsick09} have reported the detection of Fe line in the hard state of 
GX 339-4, which is seen at low luminosities of $\sim$0.14$\%$ $L_{Edd}$ using Suzaku and RXTE
observations. The truncated disk scenario is supported by their results wherein the inner 
disk radius is shown to increase by a factor of $>$27 as compared to the value found
when the source was bright. Although, the inner disk radius is shown to be dependent
on the inclination angle (i.e. $R_{in} >$35 $R_g$ at $i$=0$^\circ$ and $R_{in} >$175 
$R_g$ at $i$=30$^\circ$), the results provide a direct evidence for the absence	of the 
inner disk at low luminosities.
The detection of Fe line in the hard state of GX 339-4 is reported by \cite{shidatsu11}
and an inner disk radius of $\sim$13.3 $R_g$ is estimated by modeling its profile. 
Their results indicate that the accretion disk evolves inward as luminosity increases
in the range $\sim$0.001 $< L_X/L_{Edd} < \sim $0.02 when the source remains in 
the hard state.
The hints of disk recession can be found in \cite{petrucci14}, wherein the spectral
results of five Suzaku observations taken during the decline of 2010-2011 outburst are presented.
An inner disk radius of $<$10-30 $R_g$ is found in the first two observations, however
it remains unconstrained for the latter observations due to low statistics.
\cite{plant14} have presented a very detailed study of reflection component observed
in the joint spectral fitting of PCA and HEXTE spectra for the three outbursts of GX 339-4.
Their results support the truncation of the inner disk during the hard state, and
a decrease in the coronal height in the soft state. Truncation of disk in GX 339-4 in the 
hard state is again corroborated by \cite{plant15} with the spectral studies 
performed using XMM-Newton and Suzaku observations.
\cite{kole14} studied GX 339-4 with XMM-Newton data in its hard state, and used 
both Fe line emission and disc continuum methods to measure the 
inner radius of the accretion disk. They estimated the effects of instrumental and modeling 
uncertainties and showed that both the methods provide results consistent with the 
truncated disc model. In addition, there are several
reports where authors reanalyzed the results suggesting violation of truncation of disk
and reclaimed the finding of truncated disk.
A reanalysis of BeppoSAX observation of GX 339-4 by \cite{donegier06} confirmed
the finding of a broad Fe line caused by the extreme relativistic effects previously 
reported by \cite{miller02} and \cite{miniutti04}. However, the authors
reinstated the truncation of the inner disk by showing that the relativistic 
smearing can be significantly reduced by considering
resonance Fe K line absorption from an outflowing disk wind.
\cite{donetrigo10} reanalyzed XMM-Newton data of GX 339-4 studied by \cite{miller06-653}
and \cite{reis08} which claimed the detection of broad Fe line in the hard state. A detailed
reanalysis showed that MOS data of XMM-Newton is heavily piled-up and a broad Fe line is an 
artifact of the same. The spectrum extracted with PN timing mode data of the same observation
revealed a narrow line consistent with the truncation of disk in hard state. 
Other reports showing the truncation of accretion disk in the hard state 
for various black hole binaries include \cite{basak16}, \cite{rao15},
\cite{yuannarayan14}, \cite{cabanac09}, and \cite{gierlinski08}.

The Fe line emission is just one feature of the reflection component detected
in black hole binaries. Another major feature of this component is the broad
hump that appear at $\sim$10-30 keV. The reflection hump appearing towards 
higher energies, if not modeled properly, can give rise to an artificial 
hardening to an intrinsically soft spectrum. Therefore, it may be possible
that reflection can be the reason behind hardening of an originally soft
spectrum. It implies that the system
is not seed photon starved and hence the finding
of a broad Fe line can also be reconciled. In order to test the hypothesis, we 
have studied the black hole binary GX 339-4, which is already an object in the
debate of truncation of accretion disk in hard state \citep[e.g.][]{plant14}.
It is a stellar mass galactic black hole binary
harboring a low-mass donor and a confirmed black hole with a mass of  
5.8 $\pm$ 0.5 $M_\circ$ \citep{hynes03} and distance of $>$6 kpc \citep{hynes04}.
GX 339-4 has shown multiple outbursts in the past, which have been 
regularly monitored with RXTE. 
We have studied 4 outbursts of the object during 2002-2011 and in particularly its
spectra.
Our motivation is to fit the spectra with and without reflection component, in order
to understand the effect the component has on the photon index. 
In the next section we will
discuss the observations and the results found in the present work are discussed in the sections 2 and 3 respectively.

 \section{Observations and Spectral Fitting}           
\label{sect:data reduction}
This work presents a spectral analysis of pointed observations of 
the black hole binary \gx{} with Proportional Counter Array \citep[PCA;][]{jahoda96} 
onboard Rossi X-ray Timing 
Explorer (RXTE) during 2002-2011. We have studied a total of 1160 pointed
observations available on the High Energy
Astrophysics Science Analysis Archive (HEASARC) covering the four outbursts 
(2002-03, 2004-05, 2006-07, 2010-11) of the transient object. We study 
Standard-2 spectra from PCA data, 
and the spectral fitting 
was performed in the energy range of 3-20 keV using the spectral fitting package
{\tt XSPEC} version 12.8.2. All spectral parameters
are presented with a confidence interval of 90\% unless otherwise mentioned.
The spectra studied here belong to the different spectral states of the 
four outbursts, which is manifested by their variable shapes.
The spectral components have varying relative
strengths, indicating the changing geometry and evolution of the physical processes
in the system. Therefore, the spectra require different spectral models and we 
follow a scheme of spectral fitting where we begin with the simplest model of a
power law absorbed by the interstellar medium and increase the complexity of 
the model by including the thermal emission from the accretion disk, Fe K$\alpha$
emission and reflection continuum. 
All the models are listed in the Table (1) and
a description of the fitting procedure is given below. 

\begin{figure}
 \centering
  \includegraphics[width=8cm, height=12cm,angle=270]{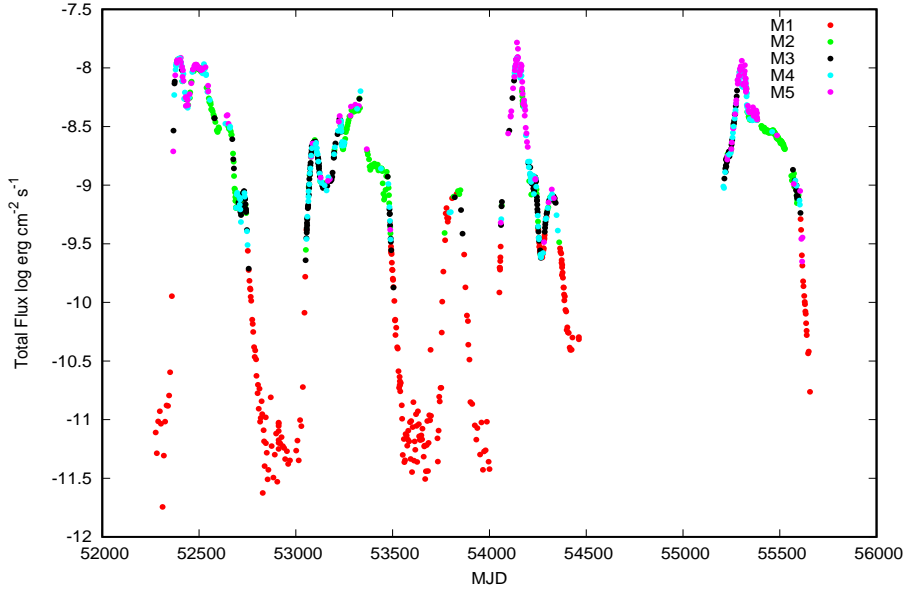}

   \caption{The plot shows the total flux in 3-20 keV energy range as a function of time for 
   the four outbursts during 2002-11. The spectra providing good fit with model M1, M2, M3, M4 
   and, M5 are shown with red, green, black, cyan, and pink respectively.} 
   \label{fluxtime}
   \end{figure}

We begin the spectral fitting of all the spectra with the simplest model 
of an absorbed power law listed as M1 in the Table (1) and 
mention the number of spectra explained with the model with
alphabets A, B, C and D belonging to 2002-03, 2004-05, 2006-07 and
2010-11 outbursts respectively.
The interstellar absorption
is modeled with the model {\tt wabs} \citep{morrison83}
available in {\tt XSPEC} and the 
absorption column is fixed at 3.74 x 10$^{21}$ cm$^2$.
It is found that a total of
261 ([A]66; [B]82; [C]96, [D]17)
spectra provide a good fit 
($\Delta \chi^2 <$ 1.2) with this model. Since there is no soft disk component
seen in this class of spectra, it is expected that the observations providing a good fit 
with this model belong to the hard spectral state. 
Fig~(\ref{fluxtime}) shows the 
variation of flux as a function of time and it can be seen that all the 
spectra falling in this class appear towards the lower flux values 
shown in red.

It is clear that the remaining spectra which 
are not well explained with the model M1
include other spectral components and the spectral model needs
to be modified to account for the additional components. Therefore,
the remaining spectra are fitted with the canonical model of a black hole binary
consisting of a soft thermal emission from a disk and a non-thermal 
component originating from the corona. The disk component is modeled 
with a multicolor disk black body model {\tt DISKBB} \citep{mitsuda84}
and the non-thermal component is modeled with {\tt POWERLAW}. 
It is found that 
none of the remaining spectra are explained with
the model and exhibit a larger residual close to 6 keV. Therefore, 
we include {\tt GAUSSIAN} in the model with the centroid energy fixed 
at 6.4 keV to account for the Fe line 
emission from the disk. This model {\tt WABS*(DISKBB+POWERLAW+GAUSSIAN)}
is listed as M2 in Table (1). The width of Gaussian is allowed as 
a free parameter.
A total of 
301 ([A]61; [B]81; [C]38, [D]121)
spectra are explained 
with this model giving reduced $\Delta \chi^2\leq$1.2. 
 As a next step in the spectral analysis, we allowed both the centroid energy and 
the width of Gaussian as free parameters and we name the model as M3 in Table (1).
The model resulted in
210([A]31; [B]89; [C]52; [D]48)
spectra giving reduced $\Delta \chi^2\leq$1.2. 
 Fig~(\ref{fluxtime}) shows the flux values for this model with pink color.
It was found that some of the remaining spectra showed improvement
in the spectral fitting when a narrow Gaussian line was included in the model
in addition to the broad Gaussian. 
This resulted in a model M4 {\tt WABS*(DISKBB+POWERLAW+GAUSSIAN+GAUSSIAN)}.
The model provides a good fit to 
200 ([A]51; [B]52; [C]51, [D]46)
spectra. For the spectral fitting of 
remaining spectra, the column density is allowed as a free parameter and 
this model M5 provides a good fit to 
47 ([A]16; [B]5; [C]2;[D]24)
spectra. The remaining spectra
are not explained with any of the above mentioned models resulting in higher
$\Delta \chi^2$ values. Model systematic errors were introduced  
for these spectra and systematic errors of 0.5\%, 1\%, 2\% and 3\%
allow to obtain a good fit for the 
87([A]25; [B]4; [C]28, [D]30), 33 ([A]3; [B]5; [C]14, [D]11), 15 ([C]10, [D]5 and 6([B]1; [C]4;[D]1)
spectra with model M5.
Table (1) summarizes the list of models, free parameters, 
systematic error and the number of spectra fitted. 
There are a number of free parameters in the spectral models and we focused
on the photon index and width of Fe line, in particular, for this study 
as discussed in the next section.

\begin{table}[h]
\bc
\begin{minipage}[]{125mm}
\caption[]{List of the models, free parameters and the number of spectra providing the good fit.}
\end{minipage}
\setlength{\tabcolsep}{1pt}
\small
 \begin{tabular}{c l l c }
  \hline
Model & & Free Parameters & No.of spectra\\

  \hline
$M_1$&  wabs*powerlaw                                    & $\Gamma$, N$_{pl}$& 261\\
$M_2$&  wabs(diskbb+powerlaw+gaussian)          & T$_{in}$, N$_{dbb}$, $\Gamma$, N$_{pl}$, $\sigma$, N$_{gau}$ & 301\\
$M_3$&  wabs(diskbb+powerlaw+gaussian)          & T$_{in}$, N$_{dbb}$, $\Gamma$, N$_{pl}$, E$_{gau}$, $\sigma$, N$_{gau}$ & 210\\
$M_4$&  wabs(diskbb+gaussian+powerlaw+gaussian) & T$_{in}$, N$_{dbb}$, $\Gamma$, N$_{pl}$, N$_{gau1}$, E$_{gau2}$, $\sigma_2$, N$_{gau2}$ & 200\\
$M_5$&  wabs(diskbb+gaussian+powerlaw+gaussian) & T$_{in}$, N$_{dbb}$, $\Gamma$, N$_{pl}$, E$_{gau1}$, N$_{gau1}$, E$_{gau2}$, $\sigma_2$, N$_{gau2}$, $N_H$ & 47\\
     &  ~~~~~~~~ 0.5\% SE  & & 87\\
     &  ~~~~~~~~ 1.0\% SE  & & 33\\
     &  ~~~~~~~~ 2.0\% SE  & & 15\\
     &  ~~~~~~~~ 3.0\% SE  & & 6\\
     
  \hline
  \multicolumn{2}{l}{\bf{Notes.}}\\
  \multicolumn{4}{l}{T$_{in}$ and N$_{dbb}$ are the temperature at the inner disk radius and normalization of {\tt DISKBB}.}\\
  \multicolumn{4}{l}{$\Gamma$ and N$_{pl}$ are photon index and normalization of {\tt POWERLAW}. E$_{gau}$, $\sigma$ and N$_{gau}$ are centroid energy, FWHM and}\\
  \multicolumn{4}{l}{normalization of {\tt GAUSSIAN}. The subscript `1' and `2' are used with parameters of the narrow and broad {\tt GAUSSIAN}}\\
  \multicolumn{4}{l}{respectively in models M4 and M5.}\\
  \multicolumn{4}{l}{SE stands for model systematic uncertainty.}
  \end{tabular}
\ec
\end{table}

   \section{Results and Discussion}           
\label{sect:discussion}
The scheme of spectral fitting discussed above is expected to separate
the hard spectra from the softer ones. It is expected that the 
hard spectra would appear towards the lower flux values. This is 
justified by the flux values shown in Fig~(\ref{fluxtime}), which presents the 2-10 keV flux as a function of 
time for the four outbursts. It is seen that a majority of the spectra explained with the model M1 appear towards lower
flux values, however, a few of the spectra appear towards higher flux as well. 
The spectra explained with model M2, M3, M4 and M5 are shown in green, black, cyan, and pink respectively. It is noticeable that the spectra
explained with model M4 and M5 appear towards higher flux in the plot.  

The standard model of accretion disk describe the thermal emission 
from the disk in terms of blackbody radiation emitted by the annuli of 
different radii, integrated over the inner and outer edge of the disk. 
The inner edge of the disk may extend down to the ISCO at the maximum
beyond which no stable orbits are allowed. The model, however does 
not account for the non-thermal emission commonly seen in the spectrum 
of black hole binaries. 
The overall spectrum of black hole binaries
including thermal and non-thermal components is explained in terms
of a geometrically thin and optically thick accretion disk along with 
geometrically thick and optically thin hot inner flow \citep{narayanyi95}. 
The hot inner flow intercepts
a fraction of soft seed photons from the accretion disk which is 
inverse Compton scattered. The photons escaping directly from the hot
flow give rise to the non-thermal component in the spectrum. A fraction 
of Comptonized radiation is reflected by the accretion disk resulting 
in an additional characteristic spectrum. The reflection spectrum
mainly consists of a broad continuum called Compton hump at $\sim$30 keV
and an Fe K$\alpha$ line at 6.4 keV. 

\begin{figure}
   \centering
 \includegraphics[width=8cm, height=12cm,angle=270]{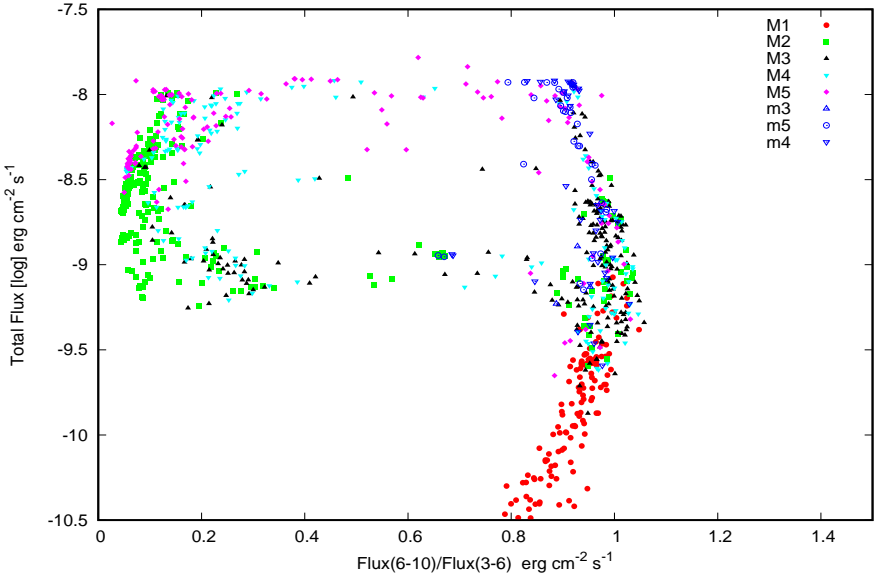}
   \caption{Hardness Intensity Diagram for 
   the four outbursts during 2002-11. 
   The spectra providing good fit with model M1, M2, M3, M4 
   and, M5 are shown with red, green, black, cyan, and pink respectively.
   The spectra with $\Gamma<1.8$ and $\sigma>1.5$ keV are shown with blue.} 
   \label{HID}
   \end{figure}

\begin{figure}
   \centering
  \includegraphics[width=8cm, height=12cm,angle=270]{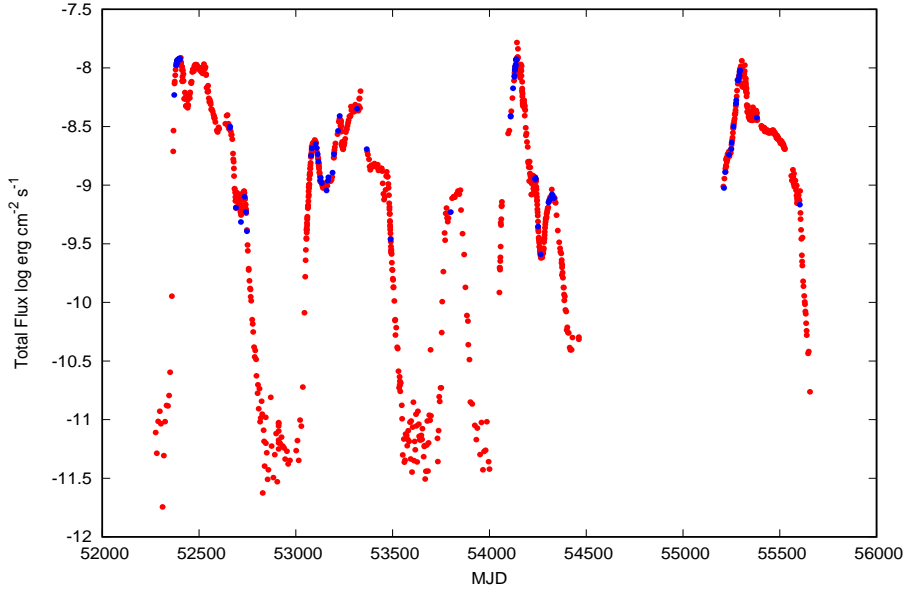}
   \caption{Total flux in 3-20 keV energy range as a function of time for 
   the four outbursts during 2002-11. The observations with $\Gamma$ $<$ 1.8 and 
   $\sigma$ $>$ 1.5 keV are shown with blue points and remaining observations with
   red. It is noticeable that the blue points appear towards the higher flux values.} 
   \label{fluxtimeredblue}
   \end{figure}

\begin{figure}
   \centering
   \includegraphics[width=8cm, height=12cm,angle=270]{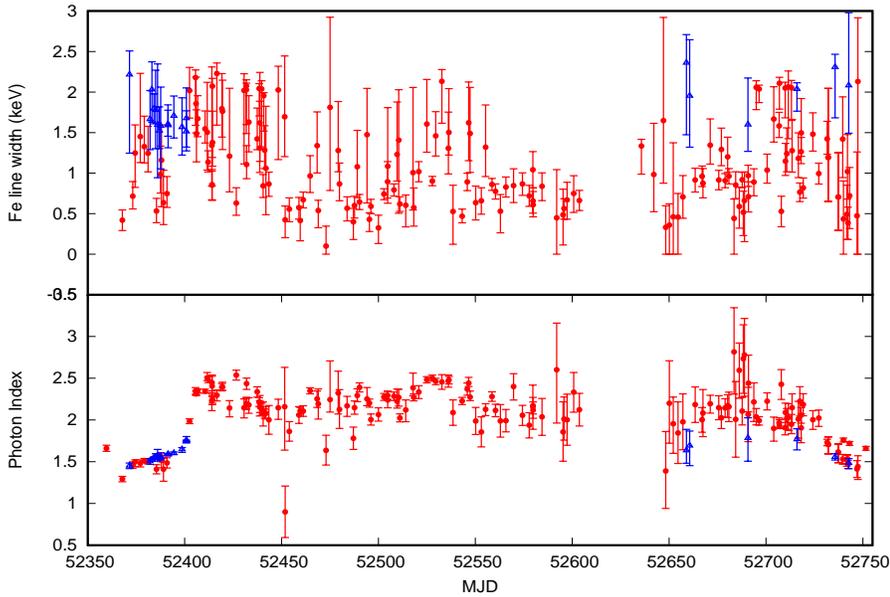}
   \caption{The plot shows the evolution of Fe line width (top panel) and photon index 
   (bottom panel) with time for 2002-03 outburst. The observations with $\Gamma$ $<$ 1.8 and 
   $\sigma$ $>$ 1.5 keV are shown with blue points and remaining observations with red.} 
   \label{sigwidthtime}
   \end{figure}

In the soft spectral state, the accretion rate is high and the disk extends
in the inner regions approaching the ISCO. The effects of general and special
relativity are prominent on the emission from the inner region resulting
in a broad and skewed Fe line. 
On the other hand, the accretion disk is expected to be truncated 
far away from the black hole
during the hard spectral state, where the effects of relativity are weaker
and a narrow Fe line is expected to be observed. Therefore, the presence
of a broad and skewed Fe line during hard state presents a violation of the 
truncated disk scenario. The disk extending close to the ISCO during hard 
state also does not explain the weaker disk component generally seen in black
hole binaries. The spectra belonging to the hard state with broader Fe line need to
be investigated in detail in order to understand the discrepancy.
We study the Hardness-Intensity diagram for the four outbursts during 2002-11.
For the sake of clarity, we represent the spectra providing good fit with model M1, M2, M3, M4 
and M5 are shown with red, green, black, cyan, and pink respectively in Fig~\ref{HID}. While 
the spectra with $\Gamma<1.8$ and $\sigma>1.5$ keV are shown by blue color with different symbols. These spectra are
fitted with models (M3, M4 and M5) are indicated as m3, m4 and m5 in Fig~\ref{HID}. The hard state spectrum is defined as a thermal Comptonization 
component with photon index  $\Gamma$ $\sim$ 1.8,  
a weak disk emission and a moderate reflection component.
We selected the spectra with photon index $<$1.8 and broad 
Fe line with FWHM $>$1.5 keV and found a total
of 76 [A:20; B:19; C:21; D:16] spectra matching these criteria.
If we interpret the broad Fe line as a result
of the relativistic effects in the vicinity of the black hole, these 
spectra present a case where the truncation of the disk stands 
violated. Therefore, we focus our further investigation
mainly on this set of spectra.
 Fig~(\ref{fluxtimeredblue}), which is the same plot as Fig~(\ref{fluxtime}) shows all 
the 76 cases marked as blue points with
almost all of them appearing towards higher flux values.
The evolution of Fe line width and photon index with time
is shown in Fig~(\ref{sigwidthtime}) for the first outburst.
The two occurrences of spectra with $\Gamma<1.8$ and $\sigma>1.5$ keV can be seen in 
Fig~(\ref{sigwidthtime}).
   
\begin{figure}
   \centering
  \includegraphics[width=13cm, height=9cm,angle=0]{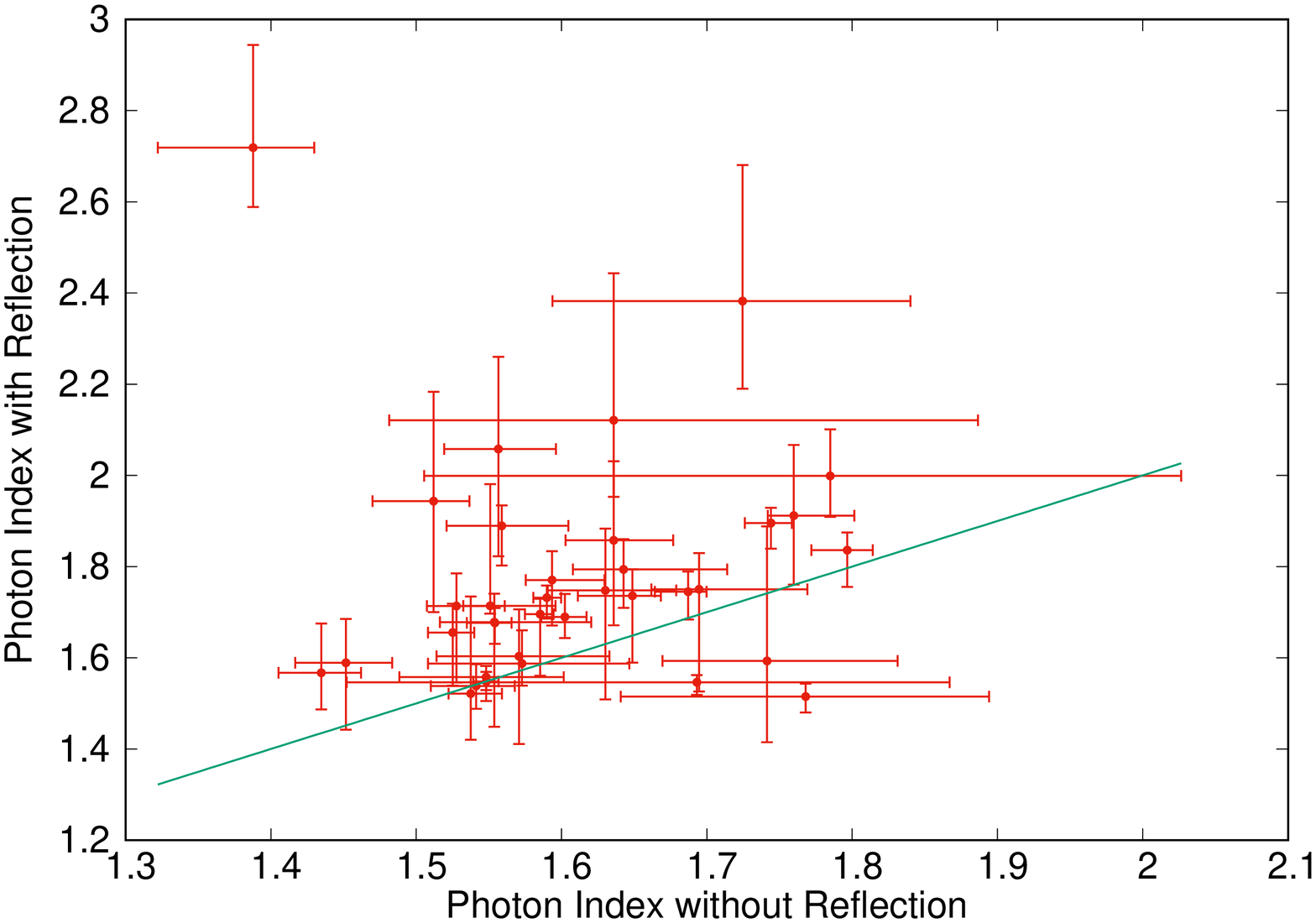}
   \caption{The values of photon index with reflection versus the photon index without 
   reflection for spectra with $\Gamma$ $<$ 1.8 and 
   $\sigma$ $>$ 1.5 keV from all the four outbursts. The plot includes the values
   of photon indices from 36 spectra where the parameter is constrained before and after the 
   addition of reflection component. The green line represent the same value of
   photon index before and after adding the reflection.} 
   \label{beforeafter}
   \end{figure}

The higher value of soft energy flux and the finding of broad Fe line
will be consistent if these spectra belong to soft spectral state with 
$\Gamma >$1.8. Therefore, a hypothesis is proposed that the observed
spectra with broad Fe line actually belong to the soft state, however
the hardening of spectra may artificially be resulted by the reflection
hump that appear towards the higher energies. Therefore, we have added
reflection component {\tt reflionx} to the 76 spectra from the four outbursts.
The change in the power law index is studied before and after
adding the reflection component in the model.
We found that the error bars of photon index are either very large or zero for the 40 spectra. 
So we have provided the spectra for the 36 observations only  where the parameter is constrained.
 Fig~(\ref{beforeafter}) shows the two values plotted against each 
other with the green line representing the same value of
power law index before and after adding the reflection.
The points lying above and below this line show the 
softening and hardening of spectra respectively. 
For some of the spectra, the photon index after adding 
reflection was found to be higher than the best-fit value obtained without
reflection component. However, the photon index with reflection component was 
found to be $>$2 only for two spectra and the remaining spectra
continued to provide photon index $<$2. 
Therefore, the occurrence of hard spectrum and broad Fe
line is not understood in the light of contribution from the 
reflection component. One of the possible solution is to model the 
spectrum with two Comptonization models \cite[see][]{yamada,basak17},
where a soft Comptonization component is considered in addition to the main 
Comptonization component. The model used by \cite{basak17}  to study 
the hard state spectrum of Cyg X-1 consists of reflection from both
Comptonization components. The soft Comptonization component models
the soft excess towards lower energies in their work. The model 
yields the best-fit with $R_{in}$=13-20 $R_g$, indicating a truncated
accretion disk in hard state.

\section{Conclusions}
The two main features of the reflection component observed in 
the spectrum of black hole binaries include an Fe emission line at 6.4 keV and 
broad Compton hump at $\sim$30 keV. 
The reflection hump appearing towards higher energies can give rise to an artificial hardening
to an intrinsically soft spectrum. 
In order to test the hypothesis, we studied spectra from the four outbursts
of \gx{} using
RXTE/PCA data observed between 2002 to 2011. We studied the spectra with 
different models of increasing complexity. 
We particularly shortlisted those cases where a broad Fe line ($\sigma >$1.5 keV)
was observed in the hard state with $\Gamma <$1.8. These spectra are refitted by 
adding reflection component and the values of the photon 
index before and after adding the reflection were compared. 
It is found that the addition of reflection component in the spectral
fitting results in the higher values of photon indices in few cases. In fact,
two spectra with photon index $<$1.8 showed photon index $>$2 
after addition of reflection component. However, there are several 
spectra where the photon index remains $\lesssim$1.8. Therefore, the results show that the 
reflection component does not completely explain the puzzle about the existence
of broad Fe line in the hard spectral state.

\bibliographystyle{raa}
\bibliography{ref1}

\label{lastpage}

\end{document}